\documentclass[amsmath,amssymb,aps,prc,twocolumn,superscriptaddress,showpacs,preprintnumbers,nofootinbib]{revtex4-1}

\def\fsize{7.5}

\usepackage{graphicx,color}
\usepackage{multirow}
\usepackage{times}
\usepackage{bm}
\usepackage{epstopdf}
\usepackage{enumerate}

\usepackage[normalem]{ulem}  
\newcommand{\comment}[1]{}
\renewcommand\sout{\bgroup \color{red} \ULdepth=-.5ex \ULset}

\begin{document}

\title{
Sensitivity of the excitation functions of collective flow
to relativistic scalar and vector meson interactions
in the relativistic quantum molecular dynamics model RQMD.RMF
}

\author{Yasushi Nara}
\affiliation{
Akita International University, Yuwa, Akita-city 010-1292, Japan}
\affiliation{Frankfurt Institute for Advanced Studies, 
D-60438 Frankfurt am Main, Germany}
\author{Horst Stoecker}
\affiliation{Frankfurt Institute for Advanced Studies, 
D-60438 Frankfurt am Main, Germany}
\affiliation{Institut f\"ur Theoretishe Physik,
 Johann Wolfgang Goethe Universit\"at, D-60438 Frankfurt am Main, Germany}
\affiliation{GSI Helmholtzzentrum f\"ur Schwerionenforschung GmbH, D-64291
Darmstadt, Germany}

\date{\today}

\begin{abstract}
Relativistic quantum molecular dynamics with scalar and vector interactions
based on the relativistic mean meson field theory, RQMD.RMF,  is developed.
It is implemented into the microscopic transport code JAM,
which includes both hadron resonances from the PDG book
and string degrees of freedom.
The sensitivity of the directed and of the elliptic proton flow
in high energy heavy-ion collisions
on the stiffness of the RMF equation of state (EoS) is examined. 
These new calculations are compared to high statistics experimental data
at mid-central Au + Au collisions in the beam energy range
$2.5 < \sqrt{s_{NN}} < 20$ GeV.
This new RQMD model with the relativistic mean field
scalar and vector meson interactions
does describe consistently, with one RMF parameter set,
the beam energy dependence of both the directed flow and the elliptic flow,
from SIS18 to AGS and RHIC BES-II energies, $\sqrt{s_{NN}}< 10$ GeV.
This is interesting, as there are different sensitivities of
these different kinds of flow to the EoS:
elliptic flow is most sensitive to the nuclear incompressibility constant,
at the moderate beam  energies $\sqrt{s_{NN}}<3$ GeV,
whereas the directed flow is most sensitive to the effective baryon mass
at saturation density at $3< \sqrt{s_{NN}}<5 $ GeV.
This self-consistent relativistic $N$-body hadronic transport approach
describes well the experimental flow data up to higher beam energies,
$\sqrt{s_{NN}} <10 $ GeV, by a stiff, monotonous EoS.
Matters abruptly change in the next higher energy range,
$\sqrt{s_{NN}}\gtrsim 10-20$ GeV:
the directed flow data show a double change of sign of the slope of $v_1$,
inverting twice in this energy range,
in sudden contradiction to the RQMD.RMF calculation
for a monotonous, stiff EoS.
This surprising oscillating behavior,
a double change of sign of the $v_1$ slope, 
points to the appearance of a hitherto unknown first-order phase transition
in excited QCD matter at high baryon densities
in mid-central Au + Au collisions.
\end{abstract}

\maketitle

\section{Introduction}

The phase structure of QCD in different regions of
the $T\mbox{-}\mu_B$ phase diagram~\cite{cbmbook}
is of fundamental interest in nuclear and astrophysics:
the structure and the maximum mass of neutron stars
and the dynamics of binary neutron star collisions,
as observed in gravitational wave detectors,
as well as the subsequent black hole formation and supernova explosions
depend sensitively on the stiffness of the nuclear equation of state (EoS)
at high temperature and density~\cite{Most:2018eaw}.
Hence, substantial effort has been devoted by theorists and experimentalists
alike to measure the EoS in the laboratory
~\cite{Stoecker:1980vf,Stoecker:1981pg,Buchwald:1984cp,%
Stoecker:1986ci,Hartnack:1994ce,
Ollitrault:1992,Danielewicz:2002pu,Stoecker:2004qu}.
New phases of hot and dense nuclear matter can be formed
in microscopic quantities in high energy heavy-ion collisions;
there, experimental data can reveal the phase properties
by unexpected changes of physical observables
when varying the beam energy (the excitation function),
the system size and the centrality of the colliding systems.

This search for a conjectured first-order phase transition
and the critical end point at high baryon density QCD matter is
a challenging goal of high energy heavy ion collision research
~\cite{DHRischke2004}.
Different species of collective flow are observed
in high energy heavy-ion collisions,
which result from a complex interplay between the initial state geometry,
nonequilibrium effects and viscosity, and the equation of state
- the pressure, which acts as a barometer for the bulk properties of
strongly interacting compressed and highly excited nuclear matter.
Anisotropic flows, such as the directed flow
$v_1=\langle\cos\phi\rangle = \langle p_x/p_T\rangle$
and the elliptic flow
$v_2=\langle\cos2\phi\rangle=\langle(p_x^2-p_y^2)/p_T^2\rangle$
are generated by the pressure already
during the early stages of the collisions.
Here $\phi$ is the azimuthal angle with respect to the reaction plane.
Distinct flow coefficients serve as sensitive messengers (barometers) of
the EoS
~\cite{Stoecker:1980vf,Stoecker:1981pg,Buchwald:1984cp,%
Stoecker:1986ci,Hartnack:1994ce,
Ollitrault:1992,Danielewicz:2002pu,Stoecker:2004qu}.
Large elliptic flow had been observed experimentally
both at the fixed target accelerators Bevalac, SIS, AGS and SPS,
as well as at the colliders RHIC and LHC.
The measured flow is in reasonably good agreement with
hydrodynamical simulations%
~\cite{Heinz:2013th,Gale:2013da,Huovinen:2013wma,Hirano:2012kj,%
Jeon:2015dfa,Jaiswal:2016hex,Romatschke:2017ejr}.

The ongoing Beam Energy Scan (BES) program~\cite{Singha:2016mna,Adam:2019wnb}
at the BNL-RHIC-STAR- and
the CERN-SPS-NA49 and -NA61/SHINE experiments~\cite{Turko:2018kvt}
seek to find the conjectured ``point of onset of deconfinement" (CPOD)
and the conjectured  associated critical point (CP).
Future experiments, such as RHIC-BESII~\cite{BESII}, STAR-FXT, CBM and HADES at
FAIR~\cite{FAIR,Sturm:2010yit}, BM@N and MPD at NICA~\cite{NICA},
HIAF at Canton, as well as the proposed J-PARC-HI~\cite{HSakoNPA2016}
will offer excellent high statistics data which will allow to
explore the highest density baryonic matter sector of QCD, and reveal the
phase structure of QCD at high baryochemical potential $\mu_B \approx 1$ GeV.

Significant progress in modeling heavy-ion collisions
at high baryon density is in order to interpret the wealth of the new data.
Improved ideal and viscous hydrodynamic theory and transport models
have been developed over the last decade,
simulating the dynamics of high energy heavy-ion collisions
\cite{Petersen:2008dd,Batyuk:2016qmb,Shen:2017bsr,Denicol:2018wdp,Akamatsu:2018olk}.
To date, all these models have not been successful in
explaining and reproducing the unique prediction of the strange looking
``double change of sign'', from a positive slope of $v_1$
to a negative slope and back to a positive slope of $v_1$,
for mid-rapidity protons in Pb + Pb collisions, which should only occur
if there is a first-order phase transition in the dense baryonic matter
(see Ref.~\cite{Stoecker:2004qu} and references therein).

These predicted negative proton $v_1$ slopes were observed
with moderate statistics more than a decade ago
at $\sqrt{s_{NN}}= 8.8$ and 17.3 GeV  by the NA49 Collaboration
in fixed target experiments at the SPS~\cite{NA49prc}.
The heavy-ion beam energy dependence of the directed flow of all hadrons,
baryons, and mesons was measured recently with much higher statistics
and in smaller steps in the BES,
from $\sqrt{s_{NN}}=7.7$ to 200 GeV,
by the STAR Collaboration at BNL's RHIC facility.
Here, the predicted drastic double change of sign of the directed flow
was now clearly discovered,
with negative mid-rapidity slopes for both,
the net proton and the net lambda directed flow
between $\sqrt{s_{NN}}= 11.5$ and 19.5 GeV~\cite{STARv1,starv1new},
in near central Au-Au collisions.
This signal can easily be distinguished
from the monotonous negative proton $v_1$ flow
predicted for peripheral collisions at high energy $\sqrt{s_{NN}}> 30$ GeV:
There, secondary interactions,
which only start after the two nuclei have passed through each other,
cause this geometrical effect,
which does not predict a double change in sign as a function of energy
\cite{Snellings:1999bt}.
All standard microscopic transport models
which do not implement a first-order phase transition
fail to reproduce the negative proton $v_1$ slope with double change of sign
at around $\sqrt{s_{NN}}=8.8-19.6$ GeV%
~\cite{Petersen:2006vm,Konchakovski:2014gda}.
In ``ideal fluid'' relativistic  hydrodynamics,
i.e., without a first-order phase transition,
there is also no change in sign of $v_1$ predicted.
However, if a first-order phase transition is put into the fluid's EoS,
it generates negative $v_1$ values at the heavy-ion energy around the
``softest point'', $\sqrt{s_{NN}}\approx 3-5$ GeV. 
This seems to be in contradiction to the positive $v_1$ data
observed at the AGS~\cite{Rischke:1995pe,Brachmann:1999xt}.
Three-fluid model (3FD) simulations \cite{Ivanov:2014ioa}
predict a minimum in the excitation function for the slope of
$v_1$ at $\sqrt{s_{NN}}\approx 6$ GeV.
Hadronic transport model calculations with strongly attractive mean fields,
supposedly simulating the effect of a first-order phase transition,
show antiflow at AGS energies~\cite{Li:1998ze}.
Microscopic transport models which take into account the effects of
the softening by the modified collision term also
predict the negative flow at $\sqrt{s_{NN}}=3-5$ GeV%
~\cite{Nara:2016phs,Nara:2016hbg}.
Hybrid models such as hydro + UrQMD~\cite{Steinheimer:2014pfa}
show no sensitivity of the directed flow on details of the EoS.

Microscopic transport models have been extensively employed to study
the dynamics of nuclear collisions (see Ref~\cite{ModelComp}
for the recent comparison of heavy-ion transport codes).
Microscopic transport models such as single-particle density
Boltzmann-Uehling-Uhlenbeck (BUU)~\cite{Bertsch:1984gb,Cassing:1990dr}
and Vlasov- Uehling-Uhlenbeck (VUU)~\cite{Kruse:1985pg}
and $N$-body quantum molecular dynamics (QMD) models~\cite{Aichelin:1986wa}
have been widely used to simulate the space-time dynamics of nuclear collisions.
These approaches used non-relativistic Skyrme forces,
where the single-particle potentials
are given by baryon density ($\rho_B$) dependent attractive and repulsive terms
and momentum dependent terms~\cite{Welke:1988zz,Li:2008gp,GiBUU}:
\begin{equation}
  V_{sk}= \alpha \rho_B + \beta \rho_B^\gamma
       + C\int d^3p'
       \frac{f(x,p')}{1+(\bm{p}-\bm{p}')^2/\Lambda^2}
  \label{eq:sky}
\end{equation}
However, these non-relativistic approaches do not reproduce
the observed beam energy dependencies of
either the directed or the elliptic flow,
if a single Skyrme parameter set is used~\cite{Danielewicz:2002pu,
E895v2,Danielewicz:1998vz,Rai:1999hz,Hillmann:2018nmd,Nara:2015ivd};
it seems that a hard EoS is required at lower beam energies
$\sqrt{s_{NN}} < 3$ GeV, and a soft EoS is favored at higher beam energies.
It was suggested that this provides evidence for a softening of the EoS,
and even for the onset of the conjectured first-order phase  transition.

A relativistic transport approach, based on the single-particle density
relativistic meson mean-field theory (RMF)
has been developed, called RBUU~\cite{RBUU,GiBUU},
RVUU~\cite{RVUU}, or RLV~\cite{RLV}.
In these covariant approaches, relativistic meson mean fields are implemented
instead of non-relativistic Skyrme potentials.
These meson fields interact with the baryons
via scalar and vector meson couplings.
This is very different from the non-relativistic potential approaches.
It was demonstrated that the beam energy dependence of
both the sidewards $\langle p_x\rangle$ and the elliptic flow are reproduced,
up to top AGS energies $\sqrt{s_{NN}}<5$ GeV, within this RBUU model,
only if the scalar and vector form factors at the vertices cut off
the interaction at high momenta ($p>1$ GeV/$c$)~\cite{Sahu:1998vz}.
A similar relativistic mean-field approach was also implemented into
the framework of the relativistic $N$-body quantum molecular dynamics
(RQMD) approach (see Ref.~\cite{Fuchs:1996uv}) and used
for simulating heavy-ion collisions up to $E_\mathrm{lab} \leq 2A$ GeV;
here only a few hadron resonances were included,
but the effects of the density dependence of the coupling constants
of the meson mean fields on the transverse flow were studied.
 
The $N$-body RQMD approach with relativistic meson mean fields
and a large number of PDG hadrons and resonances had not been developed to date.
That input is clearly most important for describing heavy-ion experiments
at higher energy: at the AGS and SPS, numerous high mass hadrons and resonances
with additional conserved charges,
like strangeness and charm, are produced in the collisions,
and multi-particle production,
efficiently described by string- and Hagedorn bag models,
plays an important role.
 
This paper presents the implementation of the relativistic  meson mean fields
into the relativistic Hamiltonian $N$-body RQMD code JAM
with a wealth of PDG book baryons, mesons, and hadronic resonances
as well as strings included.
This allows to investigate the $N$-body multi-hadron collision dynamics,
beyond the time evolution of single-particle distribution functions
like the RBUU approach.
In this paper, we specifically examine the beam energy dependence of
the directed and elliptic flow in the beam energy range
$\sqrt{s_{NN}}=2.5-20$ GeV
within the RQMD.RMF approach, where the scalar and vector interactions are
implemented in the microscopic transport code JAM~\cite{JAMorg}.

This paper is organized as follows:
Section~\ref{sec:model} describes the non-linear $\sigma$-$\omega$ model
and its implementation into the RQMD framework.
Section~\ref{sec:results} presents the results
for the beam energy dependence of the directed
and the elliptic flows, as well as the rapidity dependence of the directed flow.
The high sensitivity of the directed flow at forward-backward rapidity
on the relativistic mean-field interactions is emphasized.
The summary is given in Sec.~\ref{sec:summary}.

\section{Model}
\label{sec:model}

Here, relativistic mean-field interactions of the most abundant PDG hadrons
are implemented for the first time
into the relativistic $N$-body propagation protocol of RQMD,
as realized in the transport code JAM~\cite{JAMorg}.
The code is prepared to simulate nuclear collisions up to
$\sqrt{s_{NN}}\approx 30$ GeV.
The collision term of RQMD-JAM models particle production
by the excitations of hadrons, hadronic resonances
and strings and their decays, 
analogously as in the original RQMD code
\footnote{Here the term RQMD is used as the name of the code
developed by Sorge, Stoecker, and Greiner~\cite{RQMD1989}~\cite{Sorge:1995dp}.
However, the term RQMD is here also used for the underlying theoretical
$N$-body model, i.e., the relativistic extension of the $N$-body QMD model.
Several groups have developed RQMD codes by using approximations
to and based on the RQMD formalism by Sorge, Stoecker, and Greiner.
We use the term RQMD for the theoretical approach throughout this paper.
}
and UrQMD models~\cite{UrQMD1,UrQMD2}.
Secondary products are allowed to re-scatter, which generates
collective effects within the RQMD.RMF  approach.
Details of the collision term in JAM can be found in
Ref.~\cite{JAMorg}.

\subsection{Relativistic mean-field}

The relativistic mean-field theory employed here uses
$\sigma$ and $\omega$ meson-baryon interactions.
The corresponding equilibrium thermodynamics yields
the energy density for nuclear matter:
\begin{equation}
e = \int d^3p E^*f(p)
   + \frac{1}{2}{m_\omega^2} \omega^2
      + U(\sigma).
\end{equation}
Here $f(p)$ is the Fermi-Dirac distribution for the different baryon species.
The second term contains the $\omega$ contribution
as the zeroth component of the vector potential
$\omega^\mu$. The spatial components vanish in uniform,
stationary nuclear matter.
The single-particle energy $E^*=\sqrt{m^{*2} + p^2}$
contains the single-particle scalar potential $S$,
\begin{equation}
m^*= m - S = m - g_s\sigma
\end{equation}
through the linear coupling of the $\sigma$ meson field to the baryon.
This modifies the value of the vacuum nucleon mass $m=938$ MeV in the medium.
We consider here also Boguta's non-linear self-interactions
of the scalar field~\cite{Boguta}
\begin{equation}
  U(\sigma) = \frac{m_\sigma^2}{2}\sigma^2
         + \frac{g_2}{3}\sigma^3
         + \frac{g_3}{4}\sigma^4 \,.
\label{eq:sigmapot}
\end{equation}
The $\sigma$ field is obtained by solving the self-consistent equation
\begin{equation}
  m_\sigma^2 \sigma + g_2\sigma^2 + g_3\sigma^3=g_s\rho_s \,.
  \label{eq:sigma}
\end{equation}
Here $\rho_s=\int d^3p \frac{m^*}{E^*}f(p)$ is the scalar baryon density.
The coupling constants which reproduce nuclear matter saturation values
for a given incompressibility constant $K$
and for a given effective nucleon mass $m^*$
at ground state nuclear density are given in Table \ref{table:ns}.

\begin{table}
\caption{Parameters for the relativistic mean-field
theory with non-linear scalar interaction for a binding energy of $B=-16$ MeV
and for normal nuclear matter density of $\rho_0=0.168$ 1/fm$^3$.
A $\sigma$ mass of $m_\sigma=2.79$ 1/fm and an $\omega$ mass of $m_\omega=3.97$ 1/fm are used.
}
\begin{tabular}{ccccccc}\hline\hline
Type &$K$ & $m^*/m$ & $g_s$ & $g_v$ & $g_2$ & $g_3$ \\
     &(MeV) &   &   &  &     (1/fm) & \\
  \hline
NS1 &230 & 0.800 & 8.182  & 7.721  &  31.623   & $-3.7977$  \\ 
NS2 &380 & 0.800 & 7.211  & 7.721  & $-17.889$   &  197.64  \\ 
NS3 &380 & 0.722 & 8.562  & 9.601   & 0.4429  & 44.704 \\
\hline\hline
\end{tabular}
\label{table:ns}
\end{table}

\begin{figure}[tbh]
\includegraphics[width=\fsize cm]{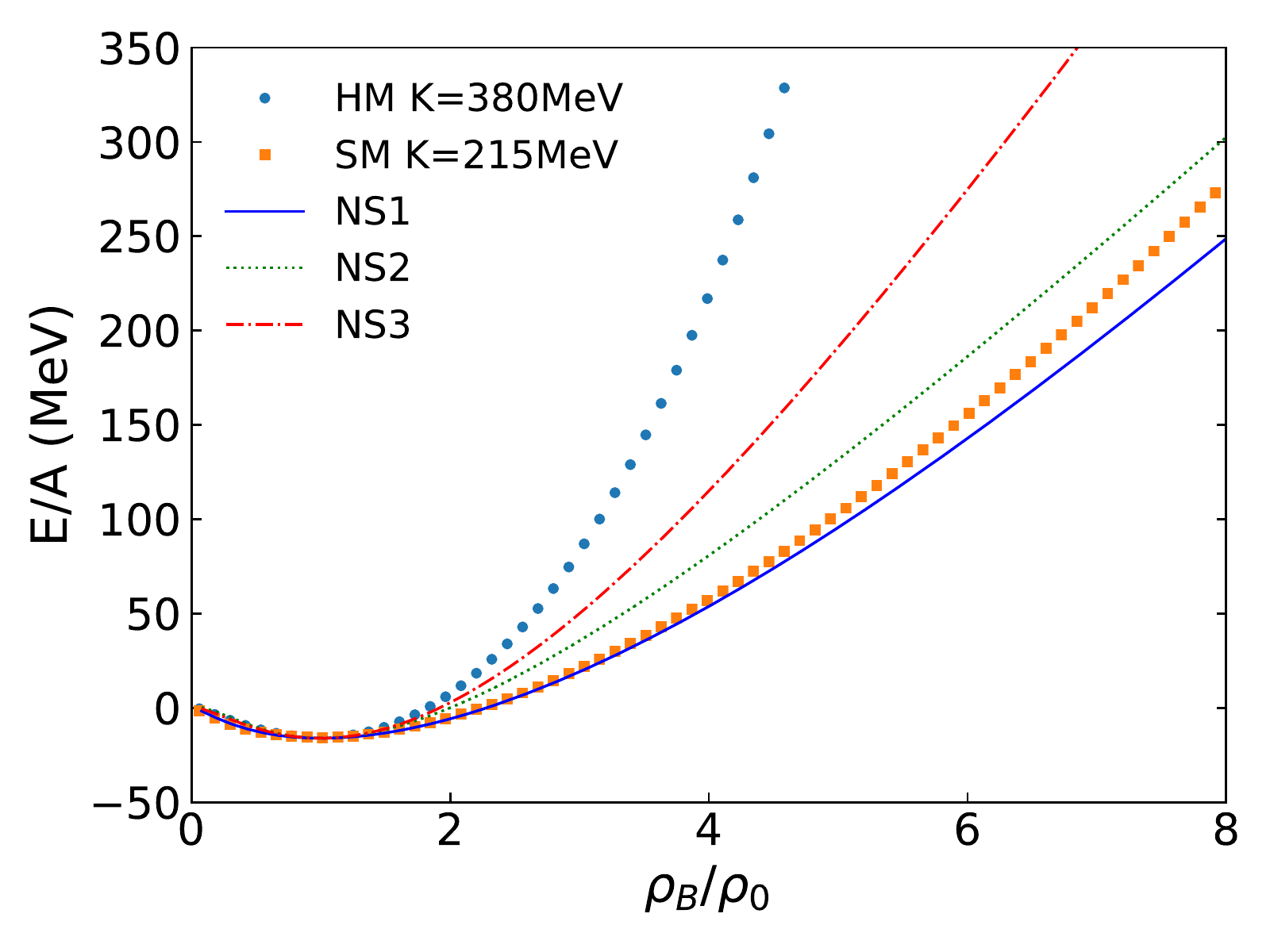}
\caption{Total energy per nucleon as a function of the normalized
baryon density at zero temperature.
}
\label{fig:eos}
\end{figure}

Figure~\ref{fig:eos} compares the energy per nucleon at zero temperature
as a function of the baryon density for different parameter sets.
Note that parameter sets with equal effective masses yield similar
EoS (see, e.g., NS1 and NS2).
Also note that smaller effective masses always yield stiffer EoS,
i.e.,  hard repulsion at high densities,
even if the ground state incompressibility constant is not large.
For instance, NS2 ($m^*/m=0.8$) and NS3 ($m^*/m=0.722$) have
identical ground state $K=380$ MeV values, but at high density, NS2 is much softer than NS3.
This behavior is due to the fact that
the effective mass at saturation density determines the strength of the vector interaction
$C_v=g_v/m_\omega$, which is independent of the scalar interaction term
due to the Weisskopf relation:
\begin{equation}
  \sqrt{m^{*2} + p^2_F(\rho_0)} + C_v\rho_0 = m + B .
\end{equation}
Here $p_F=(3/2\pi^2\rho_0)^{2/3}$ is the Fermi momentum at saturation density,
and $B=-16$ MeV is the ground state binding energy per particle of infinite nuclear matter.
Hence, the hardness of the EoS is nearly linearly related to the effective baryon mass
at high baryon densities. The hardness is barely sensitive to the ground state incompressibility
as was pointed out in Refs.~\cite{Boguta:1981px,Waldhauser:1987ed}.
For comparison, the $E/A$ values for the non-relativistic Skyrme potential
[Eq.(\ref{eq:sky})], for a hard EoS with $K=380$ (circles)
and a soft EoS with $K=215$ MeV (squares)~\cite{GiBUU}, are plotted
as a function of the baryon vector density.
Note that the non-relativistic Skyrme potential yields a similar $E/A$
as a function of the vector density as the soft relativistic parameter set
with $m^*/m = 0.8$ (NS1 and NS2).

\subsection{Relativistic quantum molecular dynamics}

The different relativistic EoS are now implemented into the framework of
the RQMD approach~\cite{RQMD1989},
formulated based on the constraint Hamiltonian dynamics~\cite{Komar:1978hc}
to simulate the $N$-body non-equilibrium dynamics.

$8N$ four-vectors $q_i^\mu$ and $p_i^\mu$
$(i=1,\dots,N)$ are used throughout
the manifestly covariant formalism for the $N$-body dynamics.
In order to reduce the number of dimensions
from $8N$ to the physical $6N$,
$2N$ constraints are employed, namely,
\begin{equation}
  \phi_i \approx 0,~~(i=1,\dots, 2N),
\end{equation}
where the sign $\approx$ stands for  Dirac's weak equality:
this equality has to be satisfied on the physical $6N$ phase space.
$2N-1$ Poincar\'e invariant constraints are used,
while the $2N$th constraint determines the evolution parameter $\tau$,
which is not necessarily Poincar\'e invariant.
The Hamiltonian of the system is constructed as the linear combination
of $2N-1$ constraints
\begin{equation}
  H = \sum_{j=1}^{2N-1} u_j(\tau) \phi_j
\end{equation}
with the Lagrange multipliers $u_j(\tau)$.
The equations of motion are given by
\begin{equation}
\begin{aligned}
   \frac{dq_i}{d\tau} &=[H, q_i]
   \approx \sum_{j=1}^{2N-1}u_j\frac{\partial\phi_j}{\partial p_j},\\
   \frac{dp_i}{d\tau} &=[H, p_i]
   \approx -\sum_{j=1}^{2N-1}u_j\frac{\partial\phi_j}{\partial q_j},
   \label{eq:motion}
\end{aligned}
\end{equation}
where the Poisson brackets are defined as
\begin{equation}
   [A, B] = \sum_{k}\left(
    \frac{\partial A}{\partial p_k}
   \cdot \frac{\partial A}{\partial q_k}
  - \frac{\partial B}{\partial q_k}
   \cdot \frac{\partial B}{\partial p_k}
       \right).
\end{equation}
The constraints are conserved in time;
\begin{equation}
  \frac{d\phi_i}{d\tau}=\frac{\partial\phi_i}{\partial\tau}
    + [H, \phi_i] \approx 0.
\end{equation}
As $2N-1$ constraints do not depend explicitly on $\tau$,
the Lagrange multipliers $u_i$ are solved as
\begin{equation}
  u_i \approx -\frac{\partial\phi_{2N}}{\partial\tau}C_{2N,i}
  ~(i=1,\dots, 2N-1),
\label{eq:constraint}
\end{equation}
where $C_{ij}^{-1}=[\phi_i,\phi_j]$.
Thus, the trajectory of the coupled system of particles in $6N$ phase space
is uniquely determined by the equations of motion Eq.~(\ref{eq:motion})
together with the Lagrange multipliers Eq.~(\ref{eq:constraint}).

Here, $N$ on-mass shell conditions are imposed:
\begin{equation}
\phi_i\equiv  p_i^{*2}-m_i^{*2}
       = (p_i - V_i)^2 - (m_i -S_i)^2,~(i=1,\dots,N)
       \label{eq:onmass}
\end{equation}
for the $i$th particle, where $V^\mu_i$ and $S_i$ are
the single-particle vector and scalar potentials,
which are functions of the baryon current $J_i^\mu$
and scalar density $\rho_{si}$.
Within the RQMD approach, these densities of the $i$th particle
are influenced by all the other particles
\begin{equation}
\rho_{s,i}=\sum_{j\neq i} \frac{m^*_j}{p_j^{*0}}\rho_{ij},~~
J^\mu_i=\sum_{j\neq i}B_j v^{*\mu}_j \rho_{ij} ,
\end{equation}
here $v_j^{*\mu}=p_j^{*\mu}/p_j^{*0}$ and $B_j$ are, respectively, the velocity
and the baryon number of the $j$th
particle, while $\rho_{ij}$ is the so-called interaction density
(the overlap of density with other hadron wave-packets) given by the Gaussian in RQMD:
\begin{equation}
  \rho_{ij}=\frac{\gamma_{ij}}{(4\pi L)^{3/2}}\exp(q^2_{Tij}/4L) .
\end{equation}
$q_{Tij}$ is the center-of-mass frame distance between
the particles $i$ and $j$, and $\gamma_{ij}$ is the respective Lorentz $\gamma$-factor which
ensures the correct normalization of the Gaussians%
~\cite{Oliinychenko:2015lva}.
Throughout this work, the Gaussian width is fixed at $L=1.0$ fm$^2$.

In addition to the $N$ on-mass shell constraints of Eq.(\ref{eq:onmass}),
the time fixation constraints, which equate all time coordinates
of particles in the computational frame, follow the Maruyama model~\cite{Maruyama:1996rn}
and Ref.~\cite{Marty:2012vs}
for the rest of the $N$ constraints:
\begin{align}
   \phi_{i+N} &\equiv \hat{a}\dot (q_i - q_N),~~~(i=1,\cdots,N-1),\nonumber\\
   \phi_{2N} &\equiv \hat{a}\dot  q_N - \tau,
\end{align}
where $\hat{a}$ is a four-dimensional vector and is a unit-vector $\hat{a}=(1,\bm{0})$
in the reference frame~\cite{Maruyama:1996rn}.
Hence, $\hat{a}$ must be
changed under Lorentz transformation into other frames, to maintain the Lorentz covariance.
A convenient choice is $\hat{a}=P/\sqrt{P^2}$ with
$P=\sum^N_i p_i$.
Then the clock-time of all particles is the same in the total center-of-mass system;
it becomes the unit-vector $(1,\bm{0})$
in the center-of-mass frame~\cite{Marty:2012vs}.
By choosing those $2N$ constraints, together with the assumption that
the arguments of the potentials are replaced by the free ones, the equations of motion are obtained
for particle $i$:
\begin{align}
\bm{\dot{x}_i} & =
    \frac{\bm{p}_i^{*}}{p_i^{*0}}
                 +\sum_{j}^N\left(
                 \frac{m_j^*}{p_j^{*0}}
                 \frac{\partial m_j^*}{\partial\bm{p}_i}
                 +v^{*\mu}_j
                   \frac{\partial {V}_{j\mu}}{\partial\bm{p}_i}
                   \right), \nonumber\\
\bm{\dot{p}}_i
          &= -\sum_{j}^N\left(
                 \frac{m_j^*}{p_j^{*0}}
                 \frac{\partial m_j^*}{\partial\bm{r}_i}
                 +v^{*\mu}_j
                   \frac{\partial V_{j\mu}}{\partial\bm{r}_i}
                   \right).
\end{align}
Note that the equations of motion for non-relativistic QMD are recovered easily
by neglecting the scalar potential 
and taking only the time component of the vector potential
into account.

The actual simulations evaluate the non-linear $\sigma$ field
as well as the $\omega$ field at each space-time point
using a local density approximation~\cite{RBUU,GiBUU,RVUU};
neglecting the derivatives of the meson fields as in Eq.(\ref{eq:sigma}) for the $\sigma$-field,
the vector potential is simply proportional to the baryon current.
This approximation is widely applied to  high energy nuclear collision simulations
\cite{Li:1995qm,Larionov:2007hy,Cassing:2015owa}. 
There, the meson field radiation and retardation effects~\cite{Weber:1990qd}
were found to be on the percent level.
The present study uses identical coupling constants for all baryons.

Note that the above mean-field interactions are combined
with the collision term, which applies Monte-Carlo methods to evaluate the scattering kernel.
These collision terms treat the change of the potentials due to scattering as small
at high energies as compared to the momenta of the outgoing particles.
Hence, the treatment of the final state phase space factors
for the outgoing particles is the same as in cascade simulations. 
Changes of the effective masses are taken into account
for outgoing particles of different species.
The violation of the global and local energy conservation
is found to be less than 1\%.
A detailed discussion of the collision term treatment is found in Ref.~\cite{JAM2}.
Relativistic on- and off-shell parton scattering was recently studied
~\cite{Moreau:2019vhw} by basing the scattering processes
on explicit matrix elements squared,
which are well defined also off-shell,
hence incorporating the changes of the final-state phase space by default.

\section{Results}
\label{sec:results}

\subsection{Beam energy dependence of the flow}

\begin{figure}[t]
\includegraphics[width=\fsize cm]{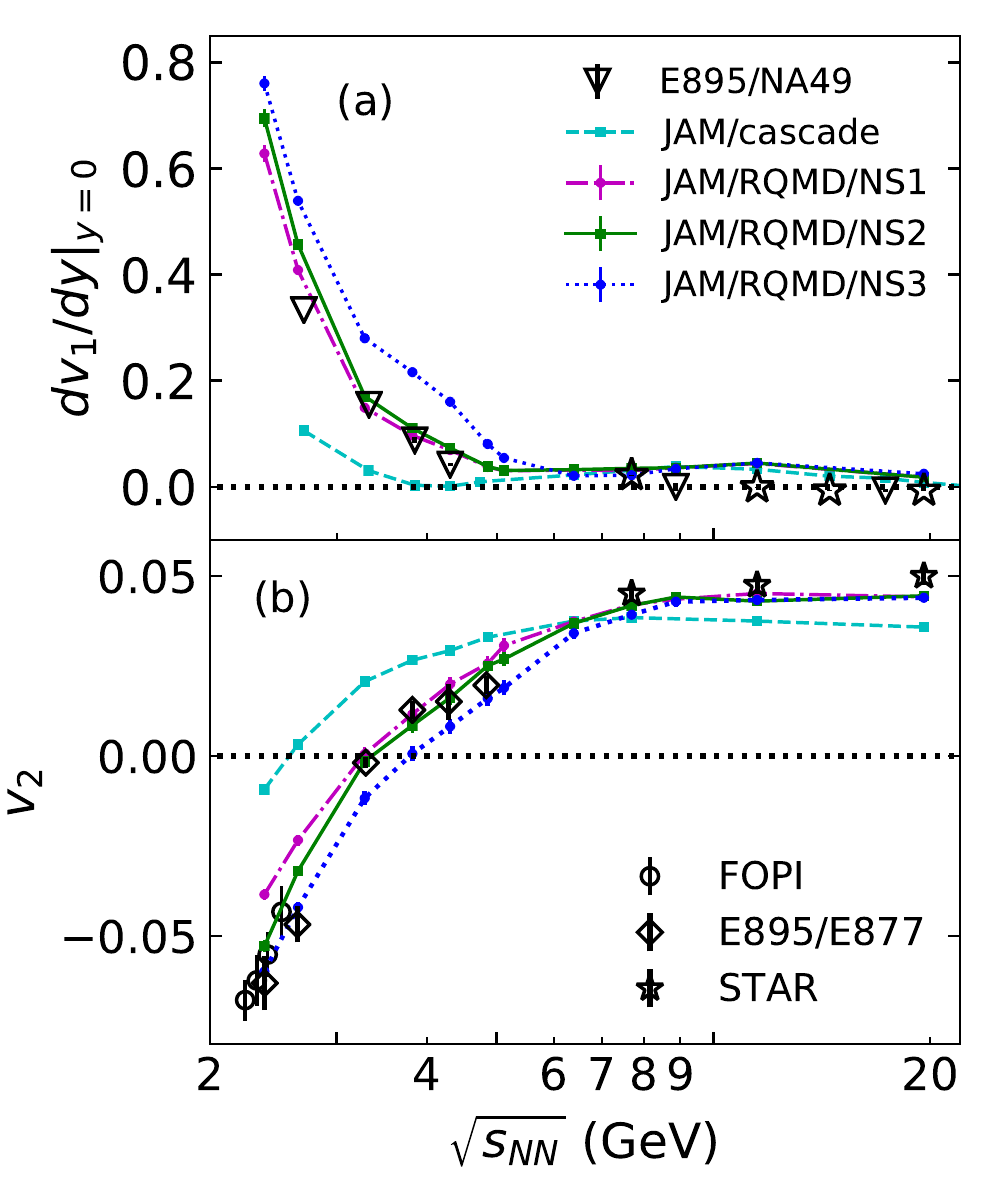}
\caption{Beam energy dependence of
(a) proton $v_1$ slope and (b) $v_2$
($|y|<0.2$)
at mid-rapidity
in mid-central Au+Au collisions ($4.6\leq b\leq9.4$ fm)
from the JAM cascade (dashed lines),
RQMD/NS1 (dash-dotted lines),
RQMD/NS2 (solid lines), and RQMD/NS3 (dotted lines).
The slopes of the proton $v_1$ are obtained by fitting
the rapidity dependence of $v_1$ to a cubic equation at $|y|<0.8$.
The experimental $v_1$ data are taken from the E895~\cite{E895v1},
NA49~\cite{NA49prl,NA49prc}, and
STAR~\cite{STARv1,starv1new,Shanmuganathan:2015qxb}
collaborations, respectively.
The STAR data~\cite{STARv2} for $v_2$ are for charged hadrons,
while the data points by E895/E877 \cite{E895v2}
and by FOPI~\cite{Andronic:2004cp} are for protons.
The JAM $v_2$ values are for protons at$\sqrt{s_{NN}}<5$ GeV,
while the higher energy data are for charged particles.
}
\label{fig:exfuncns}
\end{figure}

The upper panel of Fig.\ref{fig:exfuncns} compares
the beam energy dependence of the slope of the proton directed flow
at mid-rapidity ($|y|<0.8$) from the JAM calculations 
in the RQMD.RMF mode (JAM/RQMD)
with the E895~\cite{E895v1},
NA49~\cite{NA49prl,NA49prc}, and
STAR~\cite{STARv1,starv1new,Shanmuganathan:2015qxb} experimental data.
Only for the parameter sets NS1 ($K=230$ MeV) and NS2 ($K=380$ MeV) is
good agreement with the data found, up to $\sqrt{s_{NN}}=7.7$ GeV,
nicely demonstrating the insensitivity of the directed flow
to the incompressibility constant
at the ground state density $\rho_0$.
The fact that the NS3 parameter set simulations significantly
overestimate the $v_1$ data clearly demonstrates the strong dependence of
the directed flow on the baryons' effective mass.
The directed flow is - up to maximum  AGS- / SIS100- energies -
not sensitive to the incompressibility constant at the ground state density,
but shows a strong sensitivity to the effective mass at saturation density.
A strong influence of the effective mass on the directed flow
at the lower BEVALAC/SIS18 energies
($E_\mathrm{lab}=1-2A$ GeV) had been found in both,
RBUU and RVUU calculations~\cite{RBUU,RVUU}.
The reason for this strong dependence on $m^*$ is that smaller effective masses
imply larger values of the $\omega$ meson coupling constant,
hence the stronger repulsion.
An earlier approach based on non-relativistic Skyrme potentials
with $K=210$ MeV~\cite{Danielewicz:2002pu}
seems to correspond to the parameter sets NS1 and NS2;
both give the same hardness of the EoS.

The model predicts at beam energies $\sqrt{s_{NN}}>8$ GeV still
positive $v_1$ slopes; the directed flow slope at mid-rapidity here
was insensitive to the EoS.
Because the model yields nearly the same results as the cascade model,
it indicates that mean-field effects are very weak on $v_1$
at mid-rapidity at SPS energies.
The experimental data, however, show negative $v_1$ slopes
at $\sqrt{s_{NN}}= 10 - 20$ GeV.
It has been argued that this is an indication of the softening of the EoS.
However, note that theoretical calculations
which implement a first-order phase transition
do predict negative $v_1$ slopes for protons at AGS energies,
$\sqrt{s_{NN}}<5$ GeV,
but not at SPS energies~\cite{Rischke:1995pe,Brachmann:1999xt,%
Nara:2016phs,Nara:2016hbg,Li:1998ze}.
The present results, which are for an EoS
without a high density phase transition,
do not reproduce the observed negative $v_1$ data at $\sqrt{s_{NN}} > 8$ GeV.
Is this the long-sought-after
first-order QCD-deconfinement phase transition?

The lower panel of Fig.~\ref{fig:exfuncns} compares
the beam energy dependence of the elliptic flow $v_2$ of
the present RQMD.RMF model
with experimental heavy-ion data of the FOPI~\cite{Andronic:2004cp},
E897/E877~\cite{E895v2} and the STAR~\cite{STARv2} Collaborations.
At lower beam energies ($\sqrt{s_{NN}}<10$ GeV),
the strength of the elliptic flow is the result
of the interplay between out-of-plane (squeeze-out)
and in-plane emission~\cite{Stoecker:1986ci,Sorge1997}.
Pure cascade models lack the pressure to generate the observed
strong out-of-plane emission (negative $v_2$) at low beam energies
and result in the larger (positive sign) elliptic flow
at $\sqrt{s_{NN}}<5$ GeV.
A previous work predicted
a strong enhancement of $v_2$~\cite{Chen:2017cjw,Nara:2017qcg},
as well as of $v_4$~\cite{Nara:2018ijw}
when a first-order phase transition occurs. This is
due to the suppression of the squeeze-out as a result of
the softening of the EoS~\cite{Zhang:2018wlk}.
Here, just on the contrary,
the strongly repulsive interactions suppress $v_2$.

Figure~\ref{fig:exfuncns} demonstrates that
elliptic flow is not sensitive to the ground state incompressibility constant 
anymore when $\sqrt{s_{NN}}\approx 5 $ GeV;
both the NS1 and the NS2 parameter sets
yield very similar values of $v_2$ for beam energies $\sqrt{s_{NN}}>5 $ GeV.
However, the calculations with NS1 show less squeeze-out at lower
beam energies $\sqrt{s_{NN}}<3$ GeV,
while the NS2 parameter set yields reasonable agreement with the data. 
Thus at lower energies, elliptic flow
is sensitive to the ground state incompressibility of the EoS. 

The calculations with the relativistic mean field
predict considerably larger $v_2$ values than the cascade calculations
at SPS/BESII energies.
This agreement with the $v_2$ data at the SPS is remarkable,
as the elliptic flow achieved with the Skyrme forces
does not show an enhancement of $v_2$;
the same small values as in the cascade calculations
are found with Skyrme forces
at the beam energies $\sqrt{s_{NN}}>6$ GeV~\cite{Nara:2017qcg,Isse:2005nk}.

\subsection{Rapidity dependence of the directed flow}

\begin{figure}[tbh]
\includegraphics[width=\fsize cm]{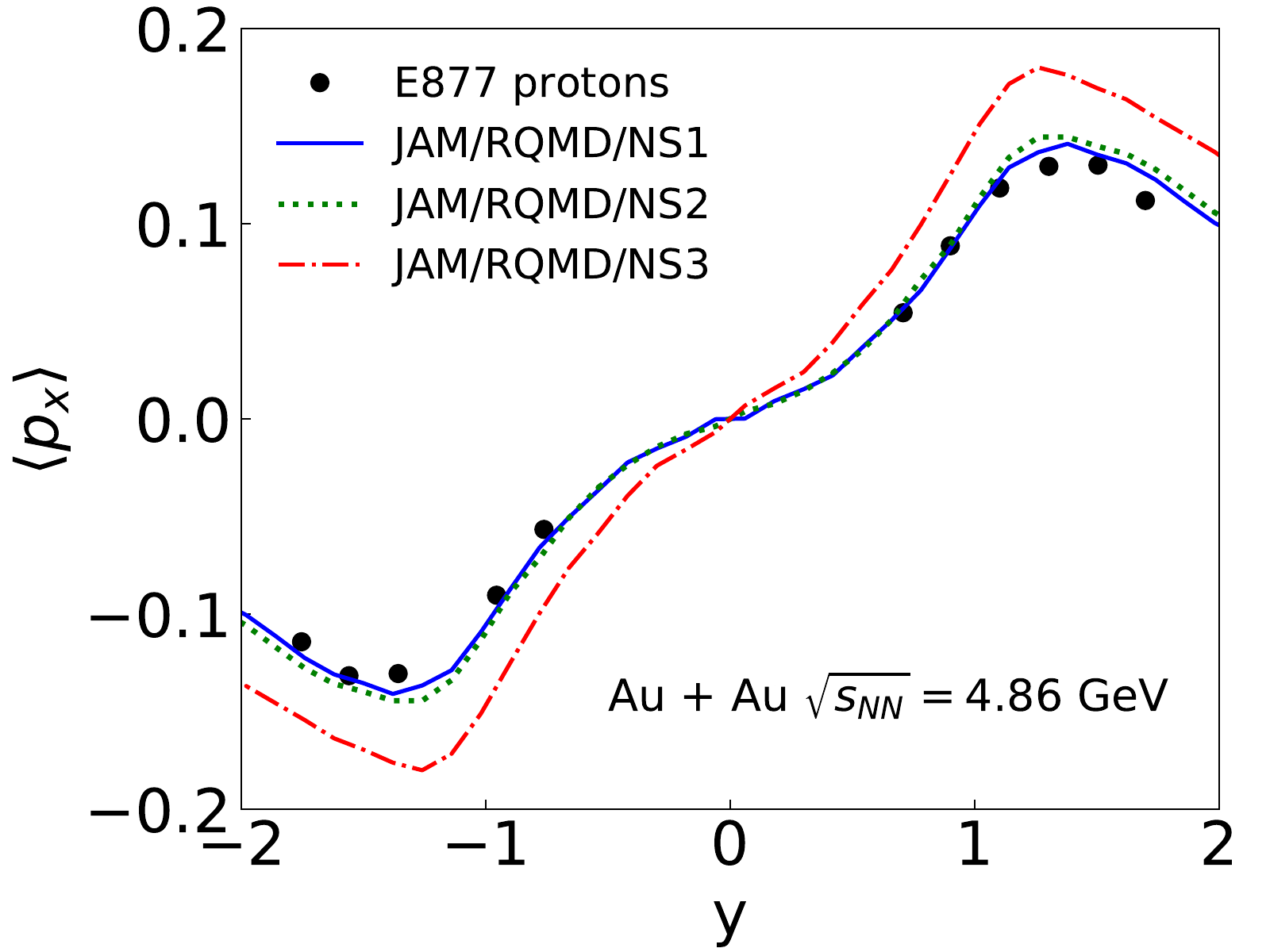}
\caption{Rapidity dependence of proton sideward flow
$\langle p_x\rangle$ in mid-central Au + Au collision
at $\sqrt{s_{NN}}=4.86$ GeV from RQMD/NS1
(solid line), RQMD/NS2 (dotted line), and RQMD/NS3 (dot-dashed line)
compared with the E877 experimental data%
~\cite{Barrette:1997pt}.  }
\label{fig:px_10gev}
\end{figure}

The rapidity dependence of the directed flow shown
in Fig.~\ref{fig:px_10gev} compares
the sidewards flow $\langle p_x\rangle$ in mid-central Au + Au collisions
at $\sqrt{s_{NN}}=4.89$ GeV, as computed from RQMD.RMF simulations
with different parameter sets, to the E877 data~\cite{Barrette:1997pt}.
The calculations using the NS3 parameter set ($m^*/m=0.722$) yield
stronger directed flow than those using the set NS2,
even though the ground state incompressibility constants
of both parameter sets NS2 and NS3 have the same value, $K=380$ MeV.
However,
the directed flow results from using the NS1 ($K=230$ MeV) parameter set
are almost identical to the NS2 results,
as it has the same effective mass parameter as NS1.
Hence, the slope of the directed flow at midrapidity
as well as its rapidity dependence is quite sensitive to
the stiffness of the high density EoS, i.e. the effective mass parameter,
but is practically insensitive to the ground state incompressibility constant
at AGS/SIS100 energies.

\begin{figure}[tbh]
\includegraphics[width=\fsize cm]{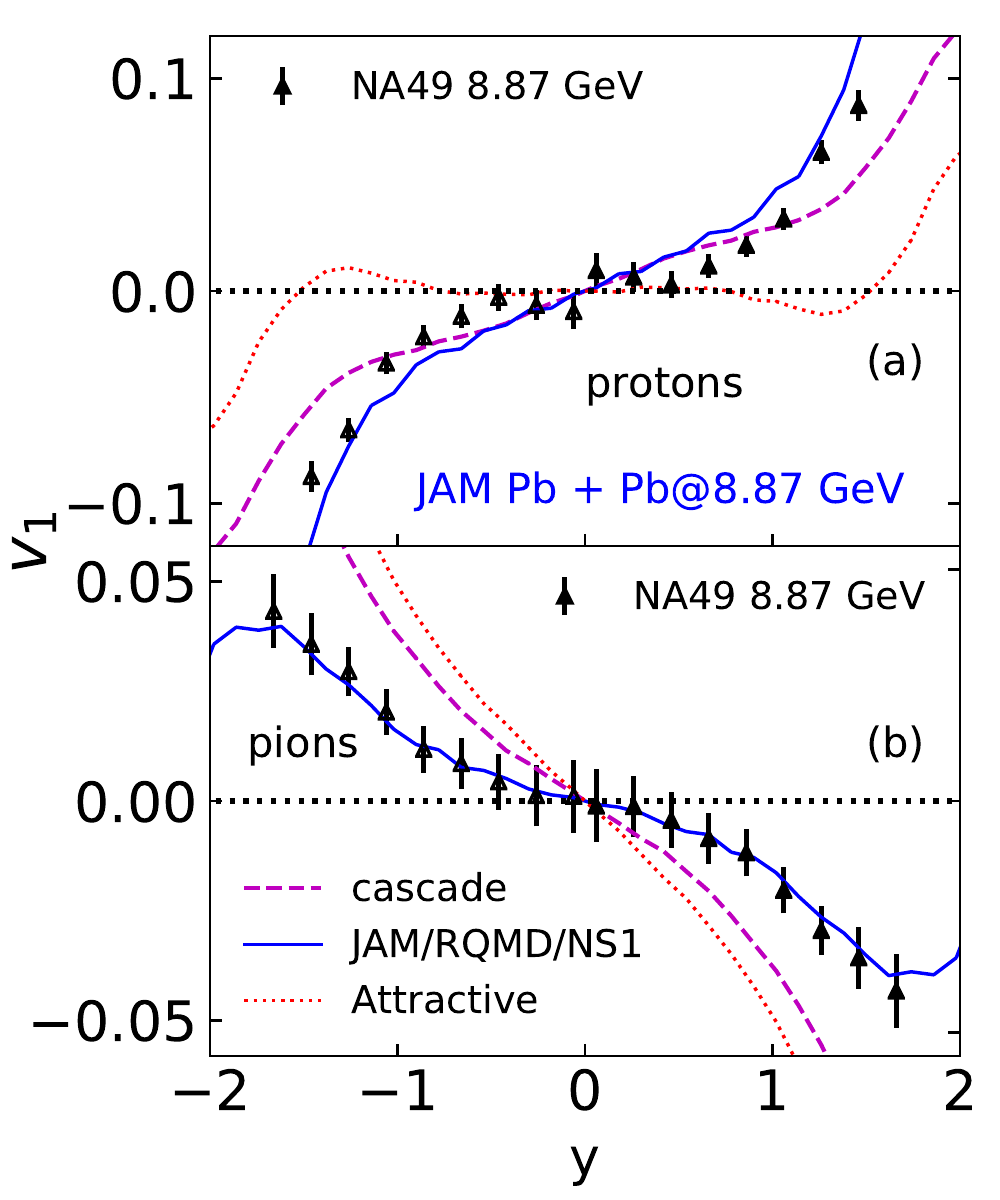}
\caption{Rapidity dependence of proton and pion $v_1$ in mid-central Pb+Pb
collision at $\sqrt{s_{NN}}=8.87$ GeV as calculated by the JAM cascade mode
(dashed line), by RQMD/NS1 (solid line),
and by JAM in the attractive scattering mode (dotted line). 
The calculations are compared to the NA49 Collaboration's experimental data~\cite{NA49prc}.
}
\label{fig:v1comp}
\end{figure}

Figure~\ref{fig:v1comp} shows the rapidity dependence of $v_1$
of protons (upper panel) and pions (lower panel)
from JAM/cascade, JAM RQMD/NS1, and JAM attractive orbit mode
in mid-central Pb + Pb collisions at 8.87 GeV together with
the NA49 data~\cite{NA49prc}.
NA49 observed the collapse of flow
at mid-rapidity
that is in good agreement with
the JAM attractive orbit mode~\cite{Nara:2016phs}.
The JAM attractive orbit mode selects only inwards scatterings
(''attractive orbit mode'')
for all binary scatterings, to mimic the softening of the EoS.
In stark contrast, calculations without a softening of the EoS, such as
JAM/cascade and RQMD/NS1, clearly  show no collapse of the directed flow at all.
Note that the RQMD/NS1 calculations show the same slope as the cascade results
at mid-rapidity ($|y|<0.5$).
This may be because baryons are not fully stopped at this energy,
and the baryon density at mid-rapidity is small, 
leading to a diminished strength of the baryon potential.
In the present calculations,
potential interactions of preformed hadrons are not included. 
They dominate the early stages of the collisions at 8.87 GeV.
This is another reason that here is no effect of the potential on the mid-rapidity directed flow,
which is most sensitive to the early compression stages of the collisions.

\begin{figure}[tbh]
\includegraphics[width=\fsize cm]{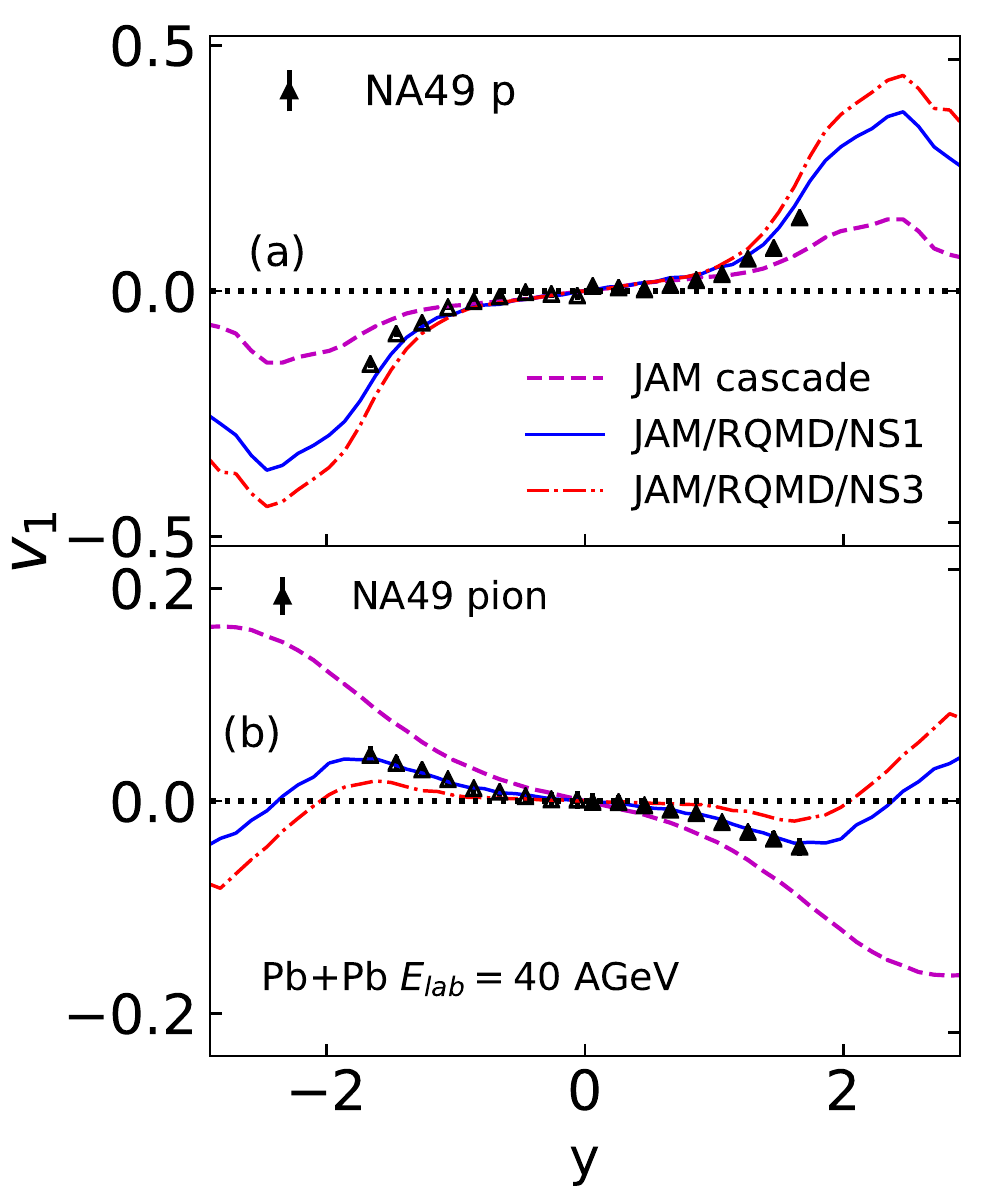}
\caption{Rapidity dependence of proton and pion $v_1$ in mid-central Pb+Pb
collision at $\sqrt{s_{NN}}=8.87$ GeV from the JAM cascade (dashed line),
RQMD/NS1 (solid line), and RQMD/NS3 (dot-dashed line)
are compared with the NA49 experimental data~\cite{NA49prc}.  }
\label{fig:v1_40gev}
\end{figure}

Figure\ref{fig:v1_40gev} compares the calculated rapidity dependence of $v_1$ of protons and
pions with the NA49 data~\cite{NA49prc} for a wider rapidity region.
$v_1$ at forward rapidity, where the baryon density is large,
does strongly depend on the EoS;
it is very sensitive to the effective mass parameter, but
insensitive to the incompressibility constant. 
At even higher beam energies, $v_1$ in forward-rapidity
regions is still sensitive to the mean field.
Hence, the $v_1$ data at forward rapidity contain valuable information about
the EoS at high baryon density in high energy heavy-ion collisions.

\section{Summary}
\label{sec:summary}

Relativistic scalar and vector meson mean-field interactions
are implemented into the JAM transport code based on
the framework of the relativistic quantum molecular dynamics model (RQMD.RMF).
The influence of the EoS
on the beam energy dependence of anisotropic flow, both directed
and elliptic flow, from SIS18 to AGS/SIS100 energies is reproduced by
the same parameter set of the scalar-vector-type interaction.
The elliptic flow at low energies $\sqrt{s_{NN}}<3$ GeV
is sensitive to the ground state incompressibility and
the directed flow is sensitive to the ground state effective mass parameter
in the EoS at energies $\sqrt{s_{NN}}<5$ GeV.
The experimentally observed negative proton directed flow at $10 < \sqrt{s_{NN}} < 20$ GeV
cannot be explained by ``normal,'' monotonously increasing
relativistic mean fields,
whereas the elliptic flow is in good agreement with the data
for the whole energy region $2.5 < \sqrt{s_{MN}}<20$ GeV.
No EoS dependent change of sign of the directed flow is 
predicted in RQMD.RMF with the ``normal'' EoS, either 
for directed or for elliptic flow,
at beam energies above $\sqrt{s_{NN}}>10 $ GeV at mid-rapidity.
This does not imply that mean field effects are negligible,
as mean-fields enhance strongly the elliptic flow at SPS energies
relative to the cascade model.
The directed flow at the forward-backward rapidity region
depends strongly on the EoS.

References~\cite{He:2016uei,Steinheimer:2018rnd,Ye:2018vbc} studied 
the influence of the mean-field potentials
on the cumulant ratios of protons  within
QMD models with Skyrme forces. 
Relativistic $\omega$ fields may suppress
the kurtosis of the baryon number distribution at high baryon densities
in mean-field nuclear matter calculations~\cite{Fukushima:2014lfa}.
An investigation of such a suppression
within the RQMD.RMF dynamical approach is under way.

The late hadronic fluid stage of collisions at RHIC/LHC energies
is usually described by hadronic cascade models.
However, the mean-field effects presented here may also be relevant
for this late evolution stage of the hadronic fluid stage
at RHIC/LHC energies.
Observables such as baryon spectra at forward rapidity
as well as flow $v_1$ and also $v_2$ can allow for a study of the mean-field
effects on the final hadronic fluid stage on an event-by-event basis.

\begin{acknowledgments}
We thank W. Cassing and J. Steinheimer
for valuable comments on the manuscript.
This work was supported in part by the
Grants-in-Aid for Scientific Research from JSPS (JP17K05448).
H. St. appreciates the generous endowment of the
  Judah M. Eisenberg Laureatus professorship.
Computational resources have been provided by GSI, Darmstadt, and LOEWE CSC, 
Goethe Universit\"at Frankfurt.
\end{acknowledgments}

\end{document}